# Near Capacity Irregular Turbo Code

**Abiodun Sholiyi[1], Jafar A. Alzubi[2*], Omar A. Alzubi[2], Omar Almomani[3] and Tim O'Farrell[4]**

[1]Shaw Communications Inc., Calgary, Canada; Abiodun.Sholiyi@Shaw.ca
[2]Al-Balqa Applied University, Al-Salt, Jordan; j.zubi@bau.edu.jo, o.zubi@bau.edu.jo
[3]The World Islamic Sciences and Education University, Amman, Jordan; Omar.almomani@wise.edu
[4]The University of Sheffield, Sheffield, UK; t.ofarrell@sheffield.ac.uk

## Abstract

The purpose of this study is to construct a near capacity Irregular-Turbo Code (I-TC) and to evaluate its performance over Gaussian channel. The methodology used to evaluate and measure the performance of the new design is by simulating the system by developing a software platform using MATLAB. The simulation carried out by implementing the system over different modulation schemes, different frame sizes, and different code rates in order to achieve a fair comparison between irregular turbo code and regular Turbo Code (TC). The simulation results showed that the Irregular Turbo Code with 64QAM modulation is capable of achieving a coding gain of 1.29 dB over its corresponding Turbo Code for a low Bit Error Rate (BER) $10^{-5}$, when used in a Gaussian noisy channel. Also, the Irregular Turbo Code designed in this paper, utilizes a single recursive Convolution encoder with short frame size of 5012 bits. The new designed codec can play an important role in many commercial applications such as Third-Generation (3G) wireless phones, Digital Video Broadcasting (DVB) systems, or Wireless Metropolitan Area Networks (WMAN), etc.

## 1. Introduction

The introduction of iterative decoding (Turbo codes) by[1] has significantly reduced the transmit power to achieve negligible Bit Error Rate (BER) with moderate to large complexity in digital wireless communication systems. In[1], Turbo Codes (TC) with a large frame size (216 bits) in an AWGN channel at code rate $\frac{1}{2}$ have been shown to converge to a low probability of error at an SNR of 0.7 dB from the Shannon limit. Short frame size (212 bits) TC at the same code rate has also been shown to be at about 1.4 dB from the Shannon limit in an AWGN channel[2,3].

In[4], Valenti used various frame sizes ranging from 40 bits to 5114 bits in a Universal Mobile Telecommunication System (UMTS) TC and showed that the larger the frame size in a TC, the closer the TC gets to the Shannon bound. In general, these results show that performance close to the Shannon bound requires large frame sizes thereby increasing complexity and latency in the system.

The unequal protection of information bits has recently been of interest, from the design of an irregular LDPC codes to the design of an Irregular Turbo Code (I-TC)[5]. Significant performance benefits stem from the extra protection on some of the bits in the frame. An I-TC was first proposed in[6], where a coding gain of 0.23 dB at a BER of $10^{-4}$ was demonstrated over the corresponding regular TC in an AWGN channel using BPSK modulation and a large frame size (217 bits).

In[7], an I-TC with a frame size of 104 bits gave a coding gain of about 0.1 dB at a BER of $10^{-5}$ over the equivalent regular TC for an AWGN channel and 8PSK modulation. Also, in[8,9] an I-TC of large frame size (217 bits) achieved a coding gain of 0.24 dB at a BER of $10^{-6}$ in comparison to the regular TC in AWGN with a BPSK modulation

**Author for correspondence*

scheme. The authors in[7,8] have used the same technique to develop their I-TC but have deployed it in different modulation schemes (8PSK and BPSK) leading to different coding gains. This shows that an I-TC with large frame size can achieve a higher coding gain over the regular TC using the same frame size.

In this paper, the performance of the I-TC in comparison to the regular TC is examined using different frame sizes and different modulation schemes. Also in this paper, a new I-TC is developed, capable of achieving a coding gain of 1.29 dB over the regular TC in an AWGN channel with short frame sizes. This result was achieved using a 64QAM modulation scheme. This shows that large coding gains can be achieved in an I-TC in comparison to the regular TC using short frame sizes. At the time of writing this report, there are no results for an I-TC with higher order modulations. These results were achieved using one quarter less iterations when compared to other I-TCs.

## 2. Design and Construction of an Irregular Turbo Code

A classical parallel TC has two Recursive Systematic Convolutional (RSC) components, separated by an interleaver[1] with 3 branches: the systematic bits, a first RSC (upper) component and a second RSC (lower) component. Figure 1(a) depicts an equivalent structure of a classical TC which has only one RSC component, where all the information bits have been repeated twice (without puncturing)[8,9]. $K_t$ and $P_t$ represent the information and parity vectors, respectively, while $\pi$ denotes interleaving.

The performance of the TC in[1] and the equivalent structure in Figure 1(a) has been shown to be the same with slight differences[7]. To design an irregular TC, the equivalent TC structure is used, where a small fraction of the information bits is repeated d times (called the degree) where d > 2, e.g., d = 4. The bits of higher degree are well protected because their a posteriori values include 4 extrinsic information bits instead of 2, for example. The general structure of an irregular TC is shown in Figure 1(b).

The non-uniform repetition divides the information bits into groups indexed by i = 2, 3,.... With each group having a certain number of repetitions $d_i$ where 2, 3, ........$T$ , and $T$ is the maximum number of repetitions[7].

The number of bits in a group i gives a fraction $f_i$ of the total number of information bits at the turbo encoder input. The output of the repeater is then randomly interleaved (a random vector interleaver) and passed to the RSC component of the I-TC. The parity bits from the RSC constituent and the original information bits before the non-uniform repetitions are then transmitted. The parity bits are punctured so as to target certain code rates.

In designing the I-TC in this paper, 4 different puncturing patterns were empirically developed. The puncturing pattern 101101110 (the 2nd, 5th and 9th parity bits are punctured for every 9 parity bits), 10110 (the 2nd and 5th parity bits are punctured for every 5 parity bits), 11110 (the 5th parity bit is punctured for every 5 parity bits) and the 10 (the 2nd parity bit is punctured for every 2 parity bits) have been empirically found to be effective. The following equations show the relationship between the irregular TC rate, the puncturing pattern and the various non-uniform repetition degrees.

$$\sum_{2}^{T} f_i = 1 \tag{1}$$

$$\sum_{2}^{T} i.f_i = \bar{d} \tag{2}$$

Where, $\bar{d}$ is the average bit degree.

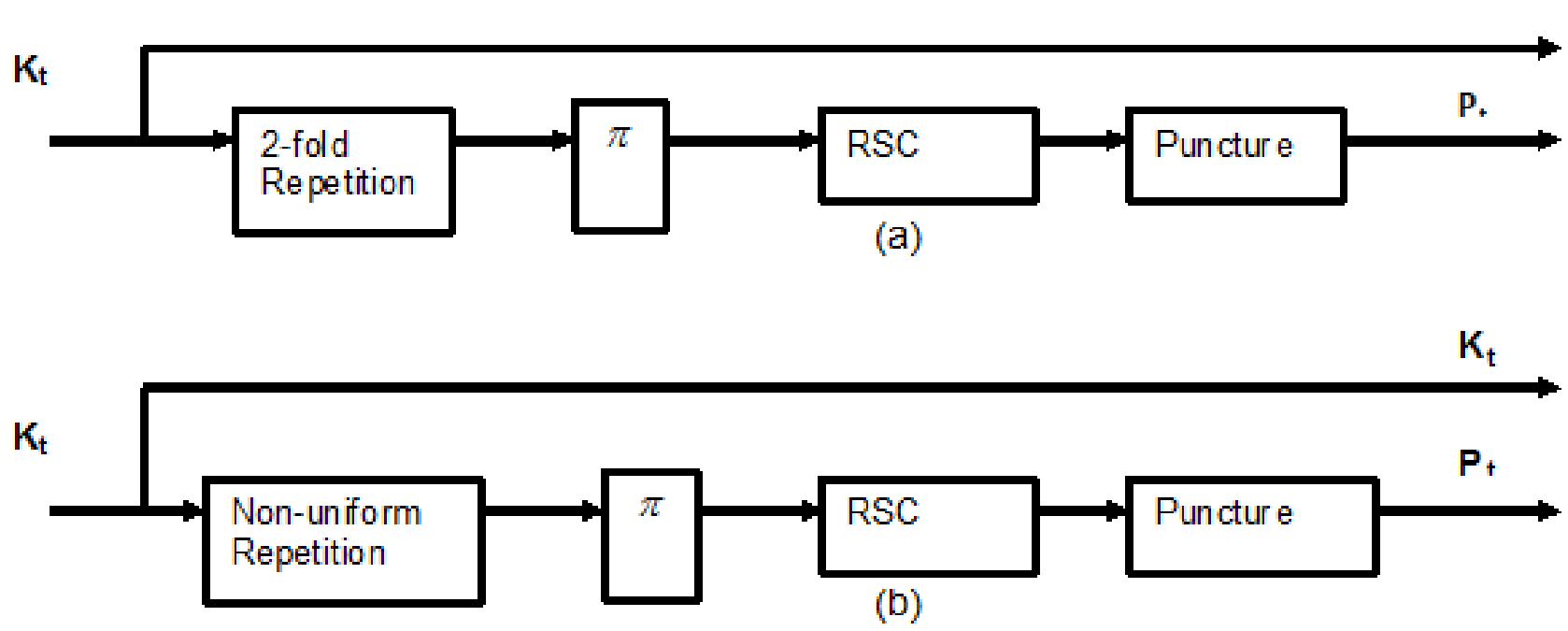

**Figure 1.** (a) Equivalent encoding structure for a regular Turbo Code and (b) Encoding structure for an Irregular Turbo Code.

$$R = \frac{1}{1+\bar{d}\left(\frac{1}{\theta}-1\right)} \quad (3)$$

Where, $\theta = \frac{1}{2-f_\theta}$ and $f_\theta$ is the fraction of the parity bits that has been punctured.

The decoding of the I-TC requires only one Log-MAP (the modified BCJR algorithm) Soft-Input Soft-Output (SISO) decoder.

Firstly, the received information bits $K_r$ are repeated (as at the transmitter) then randomly interleaved using the same interleave pattern to give $K'_r$ and then sent into the SISO decoder together with the received parity bits $P_r$ and an initial a priori value A of equal probability (i.e. zero log likelihood).

Secondly, the extrinsic information bits E at the output of the SISO decoder are computed by an extrinsic computation block in order to derive new extrinsic information for each information bit. The extrinsic information bits are de-interleaved before being transferred into the extrinsic computation block. Inside the extrinsic computational block, each information bit with degree $d_i$ will then have a new extrinsic information value which is the product (sum when using log likelihood) of the other $d_i - 1$ extrinsic information values. At the output of the extrinsic information block, the new extrinsic values are then interleaved to give the new a priori values for the next decoding iteration[10].

Figure 2 depicts the decoding process for an I-TC where π and π' denote the interleaving and deinterleaving functions, respectively. The block S in the schematic diagram is used to select the decoded bits in the format of the originally generated bits to give $K_d$ for final comparison with the originally generated bits $K_t$.

In the extrinsic computational block depicted in Figure 2, the extrinsic information at iteration for the $k^{th}$ bit $E_{jk}$ with a degree of repetition of $d_i$ is recalculated in the log domain using equation (4).

$$E_{jk} = \sum_{\substack{l=1 \\ l \neq k}}^{d_i} E_{jl} \quad (4)$$

## 3. System Parameters

The channel was modeled by computer simulation using MATLAB. A random interleaver with a frame size of 5012 bits per frame was used. The simulation was performed for a Gaussian noisy channel. The actually represented UMTS generator and feedback polynomial of the RSC were 15 and 13, respectively. UMTS Turbo Code rates used were $\frac{1}{2}$ and $\frac{1}{3}$ and the simulations were performed in the BPSK, QPSK, 16QAM and the 64QAM modulation schemes in the Gaussian noisy channel[11].

The log-MAP version of the BCJR algorithm was used as the decoder in all cases. For each simulation, a curve showing the BER versus the energy per bit to noise energy ratio ($E_b/N_0$) is graphed, where the noise energy is the single-sided noise spectral density $N_0$ of the channel[12,13].

## 4. Performance of Irregular Turbo Code in AWGN

The BER performance of the I-TC is examined using frame size of 1003, 5012, 10016 and 20072 bits with different modulation schemes in comparison to the regular TC with the same frame sizes. The simulations were done using MATLAB. The number of iterations required for the I-TC and the regular TC to converge to a low BER was also recorded.

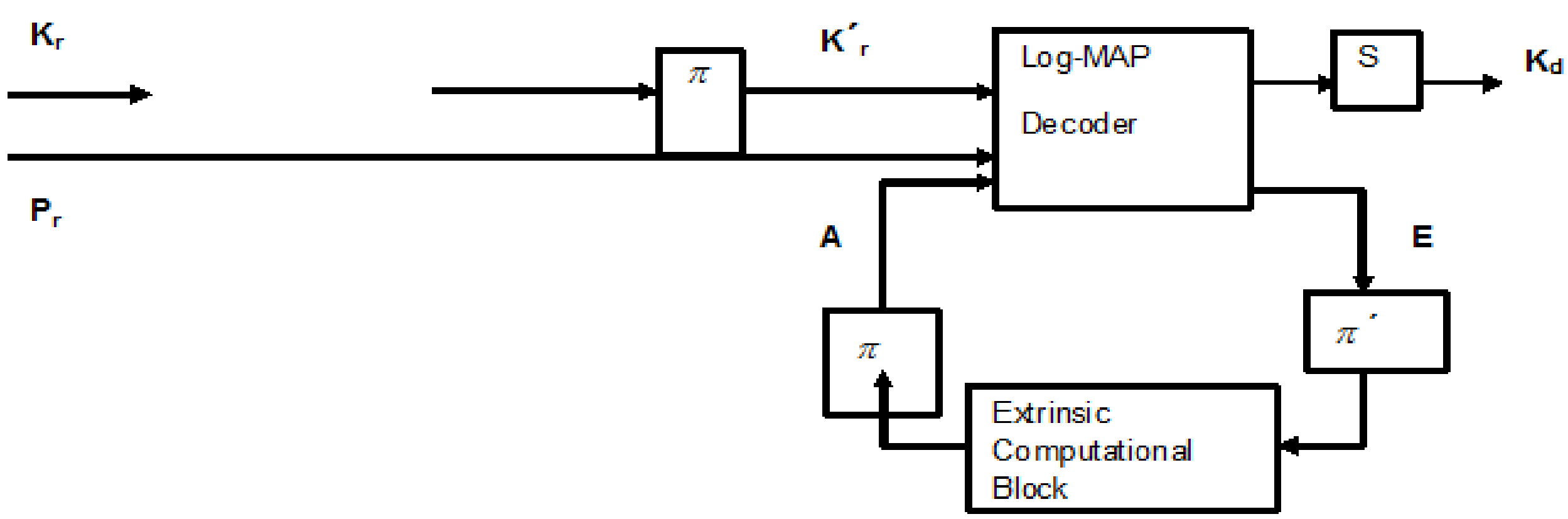

**Figure 2.** Iterative decoding structure of an Irregular Turbo Code.

Also in this section, the throughput curves for the I-TC and the regular TC have been plotted. Figures 3 to 6 show the BER performance of an I-TC in comparison to the regular TC in AWGN channel at code rate $\frac{1}{3}$ (with the exception of the QPSK I-TCs at code rate 0.40) with different modulation schemes.

Table 1 shows the puncturing patterns in the ITCs, the number of iterations required to converge to a BER of $10^{-5}$ in the I-TCs and TCs, the required $E_b/N_0$ value for convergence in the I-TCs and TCs, as well as the corresponding degree profiles used to achieve this for the I-TCs.

The complexity of the I-TC in comparison to the TC is quantified by the number of iterations required to converge to a low bit error rate, as well as the number of decoding components required to do this. It should be noted that a single iteration in the irregular TC consists of only one Log-MAP decoder unlike in the regular

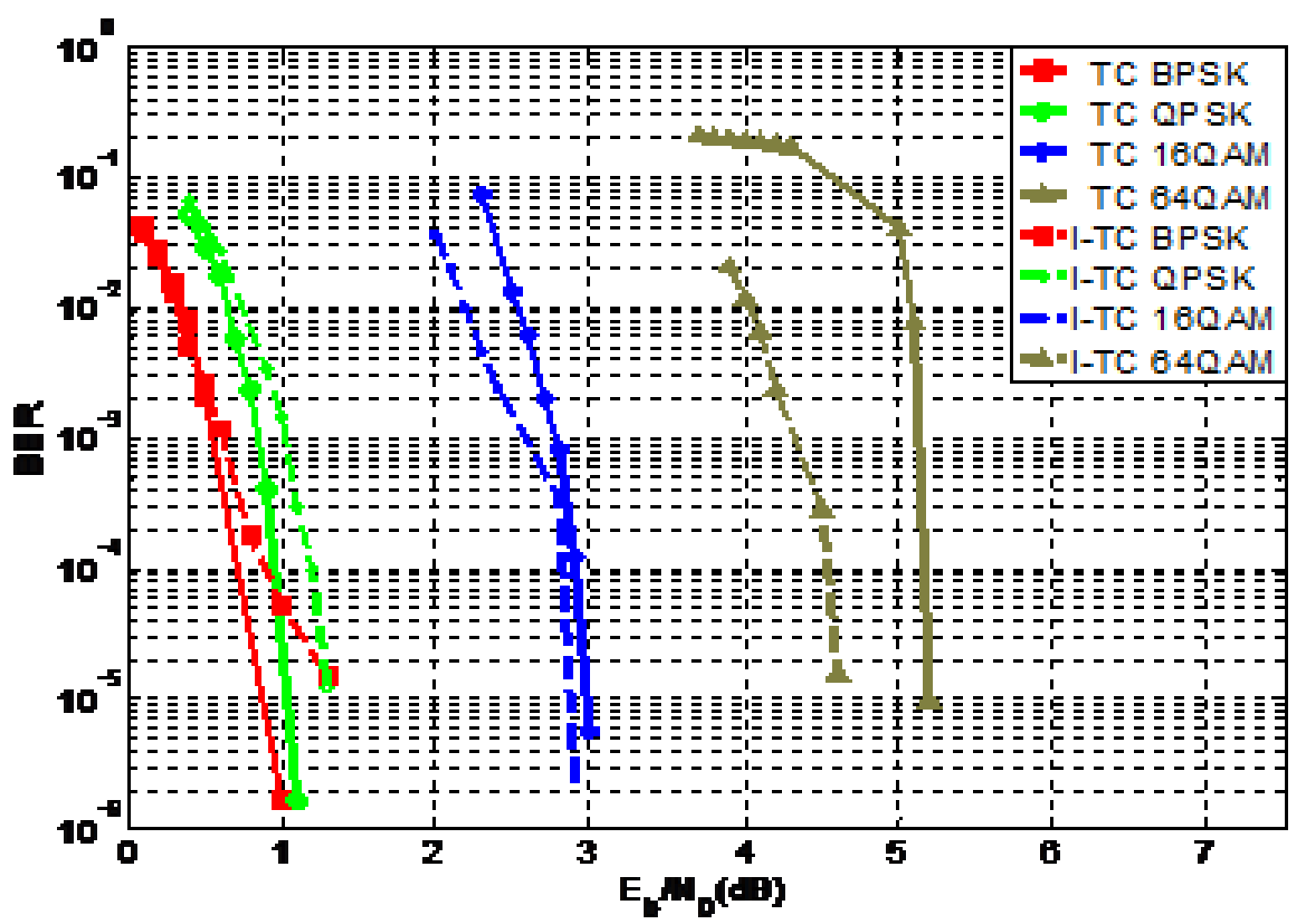

**Figure 3.** BER performance for I-TC and TC in AWGN with a frame size of 1003 bits.

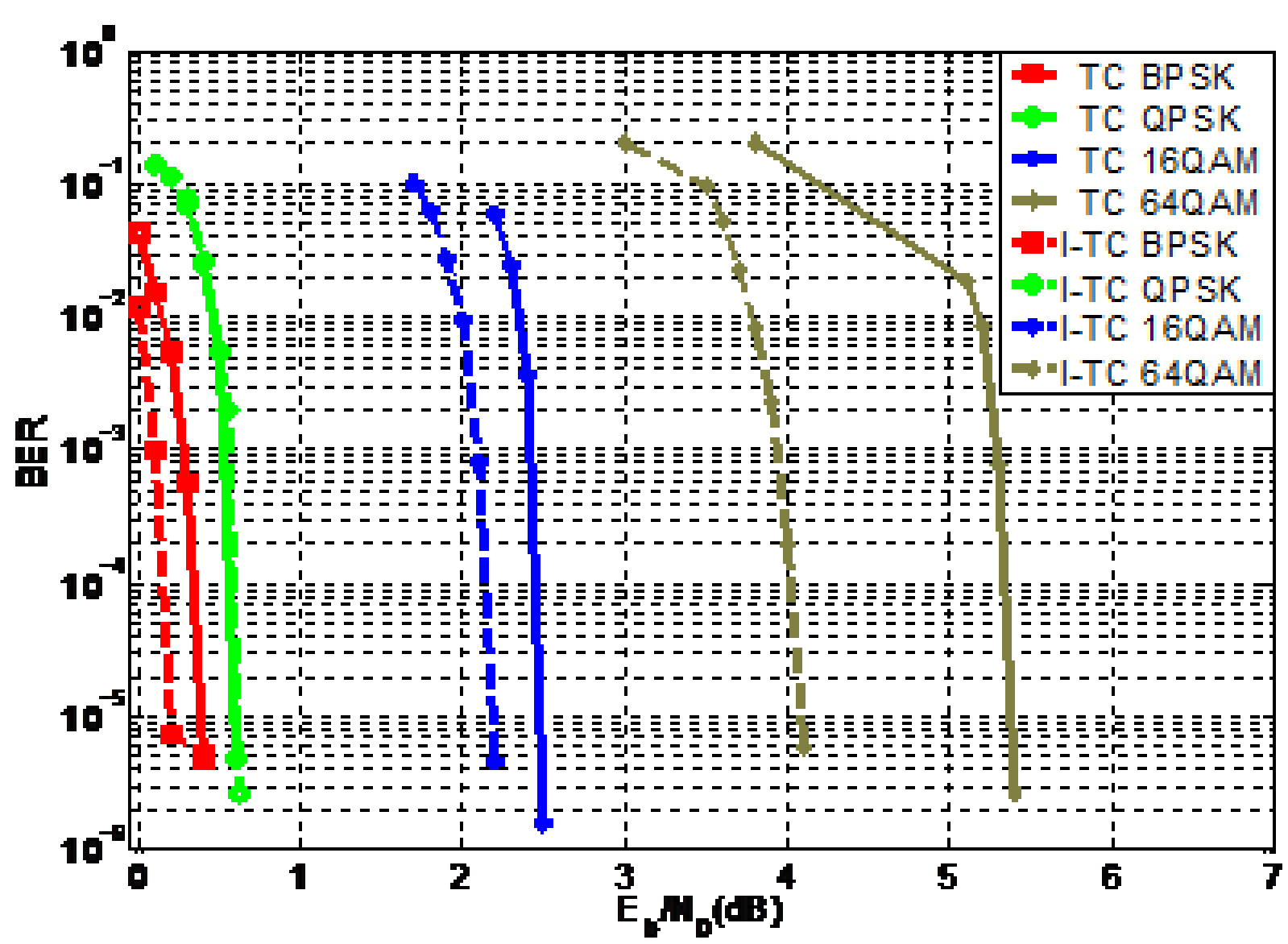

**Figure 4.** BER performance for I-TC and TC in AWGN with a frame size of 5012 bits.

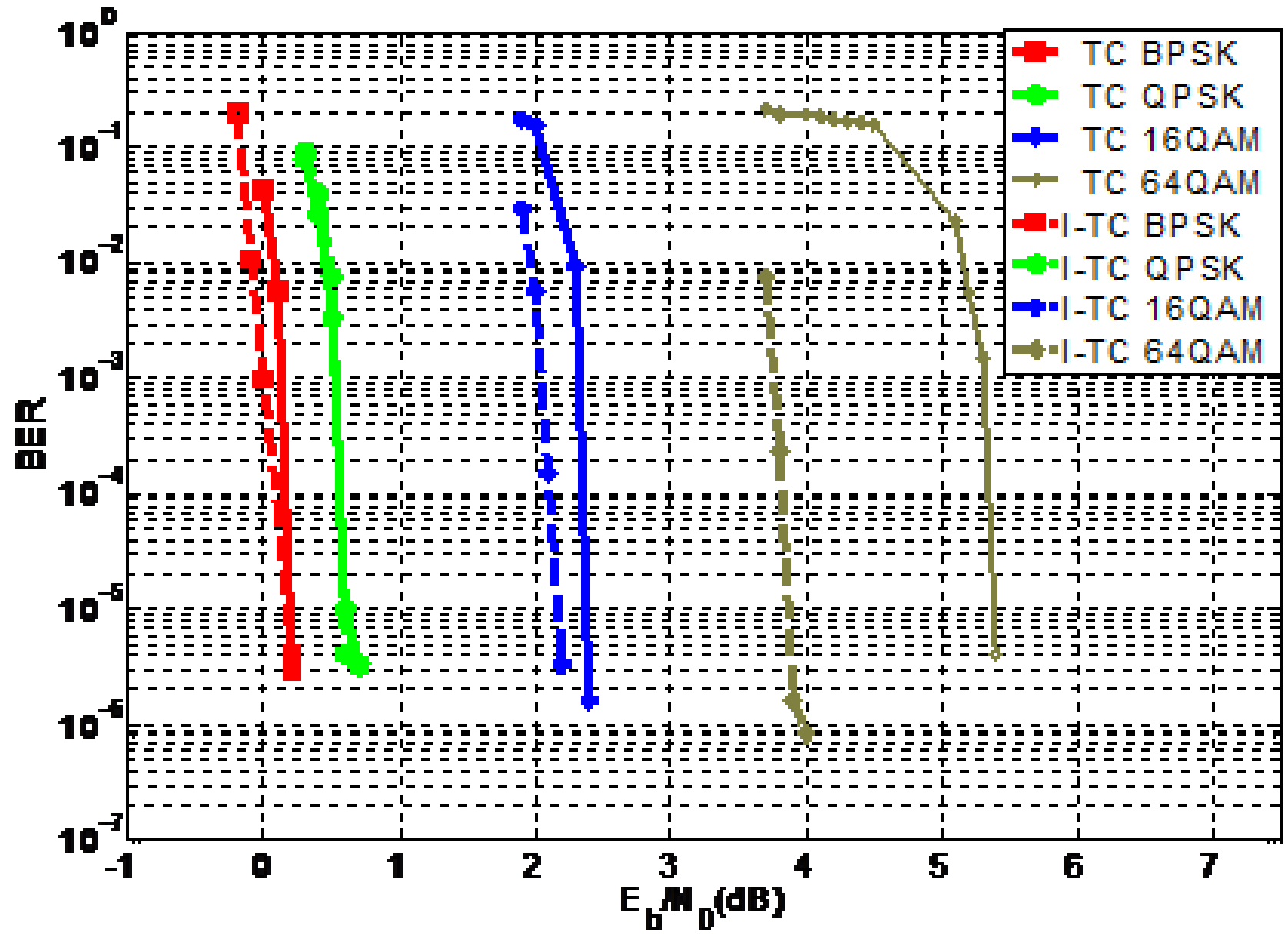

**Figure 5.** BER performance for I-TC and TC in AWGN with a frame size of 10016 bits.

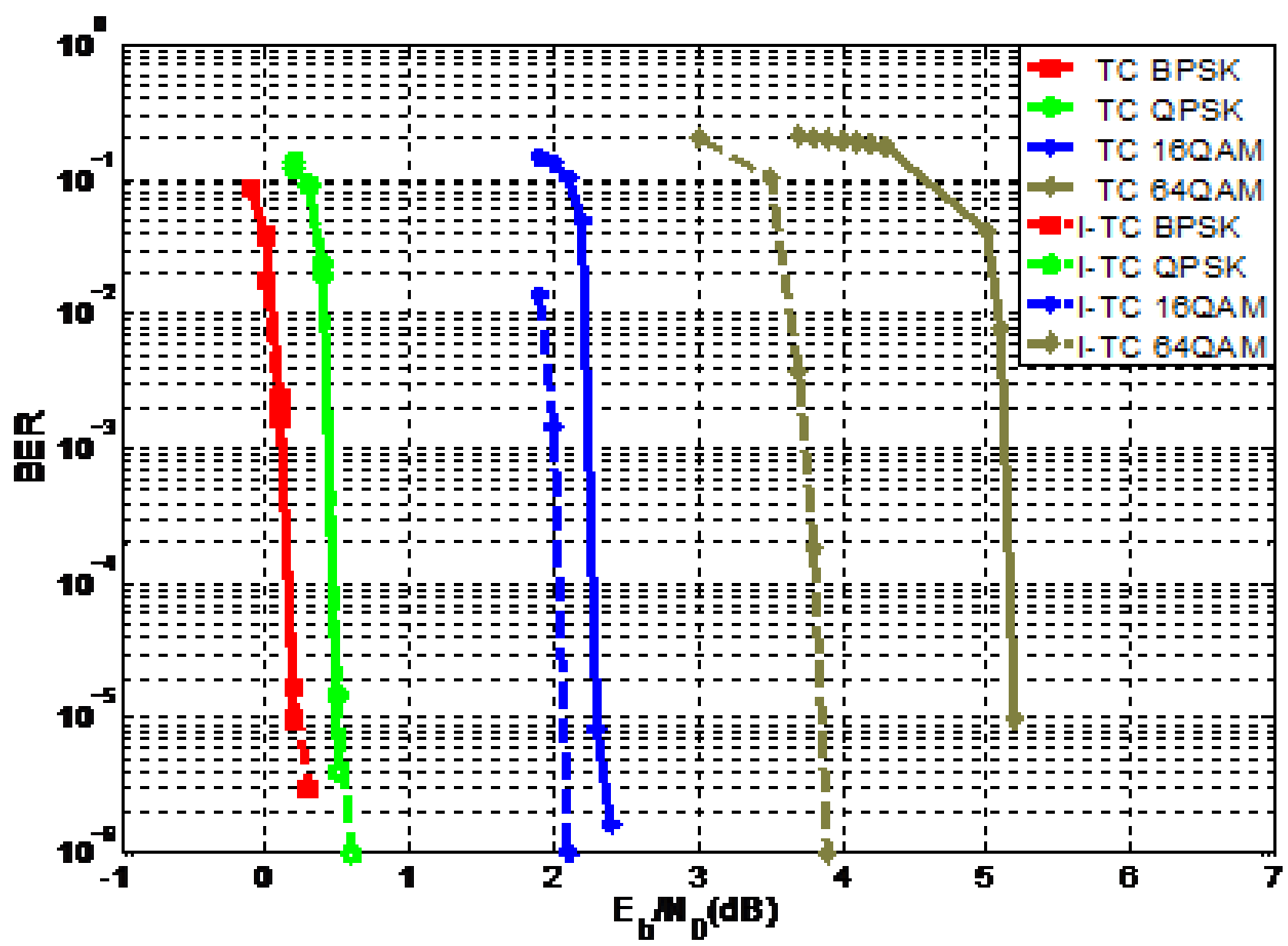

**Figure 6.** BER performance for I-TC and TC in AWGN with a frame size of 20072 bits.

TC where a single iteration consists of two Log-MAP decoders. Also previous I-TCs have used about 100 iterations to achieve a low probability of error[8,9].

Figures 3 to 6 show that, the higher the modulation order the higher the coding gain of the I-TCs over the TCs. Also, Table 1 shows that, the larger the frame size, the larger the coding gains in the I-TCs and the TCs. This shows that larger frame size codes are closer to Shannon bound than shorter frame sizes as shown in[4]. The BER performance of the I-TC in comparison to the regular TC was then evaluated with four different modulation schemes using a short frame size (5012) at code rates $\frac{1}{3}$ and $\frac{1}{2}$ (code rates 0.40 and 0.41 in the case of QPSK) as shown in Figures 7 to 10.

**Table 1.** I-TC and TC at different frame sizes with their converging $E_b/N_0$ in AWGN

| Frame Size | Code rate | Modulation Scheme | Converging $E_b/N_0$ (dB) | | Number of iterations | | Puncture pattern for I-TC | Bit degree $d_i$ (fractions $f_i$) |
|---|---|---|---|---|---|---|---|---|
| | | | TC | I-TC | TC | I-TC | | |
| 1003 | 0.33 | BPSK | 0.88 | 1.30 | 9 | 21 | 11101101110 | 2(0.888), 8(0.06),9(0.052) |
| | 0.40 | QPSK | 1.04 | 1.30 | 12 | 16 | 11101101110 | 2(0.96), 6(0.04) |
| | 0.33 | 16QAM | 3.00 | 2.85 | 14 | 19 | unpunctured | 2(0.99), 7(0.01) |
| | 0.33 | 64QAM | 5.90 | 4.60 | 22 | 24 | 11101101110 | 2(0.85), 7(0.15) |
| 5012 | 0.33 | BPSK | 0.40 | 0.20 | 11 | 31 | 11101101110 | 2(0.888), 8(0.06),9(0.052) |
| | 0.40 | QPSK | 0.70 | 0.62 | 15 | 36 | 11101101110 | 2(0.96), 6(0.04) |
| | 0.33 | 16QAM | 2.50 | 2.20 | 14 | 19 | unpunctured | 2(0.99), 7(0.01) |
| | 0.33 | 64QAM | 5.40 | 4.10 | 24 | 16 | 11101101110 | 2(0.85), 7(0.15) |
| 10016 | 0.33 | BPSK | 0.19 | 0.17 | 11 | 26 | 11101101110 | 2(0.888), 8(0.06),9(0.052) |
| | 0.40 | QPSK | 0.60 | 0.60 | 10 | 22 | 11101101110 | 2(0.96), 6(0.04) |
| | 0.33 | 16QAM | 2.30 | 2.10 | 18 | 17 | unpunctured | 2(0.99), 7(0.01) |
| | 0.33 | 64QAM | 5.38 | 3.86 | 7 | 18 | 11101101110 | 2(0.85), 7(0.15) |
| 20072 | 0.33 | BPSK | 0.19 | 0.17 | 11 | 18 | 11101101110 | 2(0.888), 8(0.06),9(0.052) |
| | 0.40 | QPSK | 0.49 | 0.52 | 10 | 30 | 11101101110 | 2(0.96), 6(0.04) |
| | 0.33 | 16QAM | 2.30 | 2.10 | 13 | 17 | unpunctured | 2(0.99), 7(0.01) |
| | 0.33 | 64QAM | 5.20 | 3.85 | 13 | 16 | 11101101110 | 2(0.85), 7(0.15) |

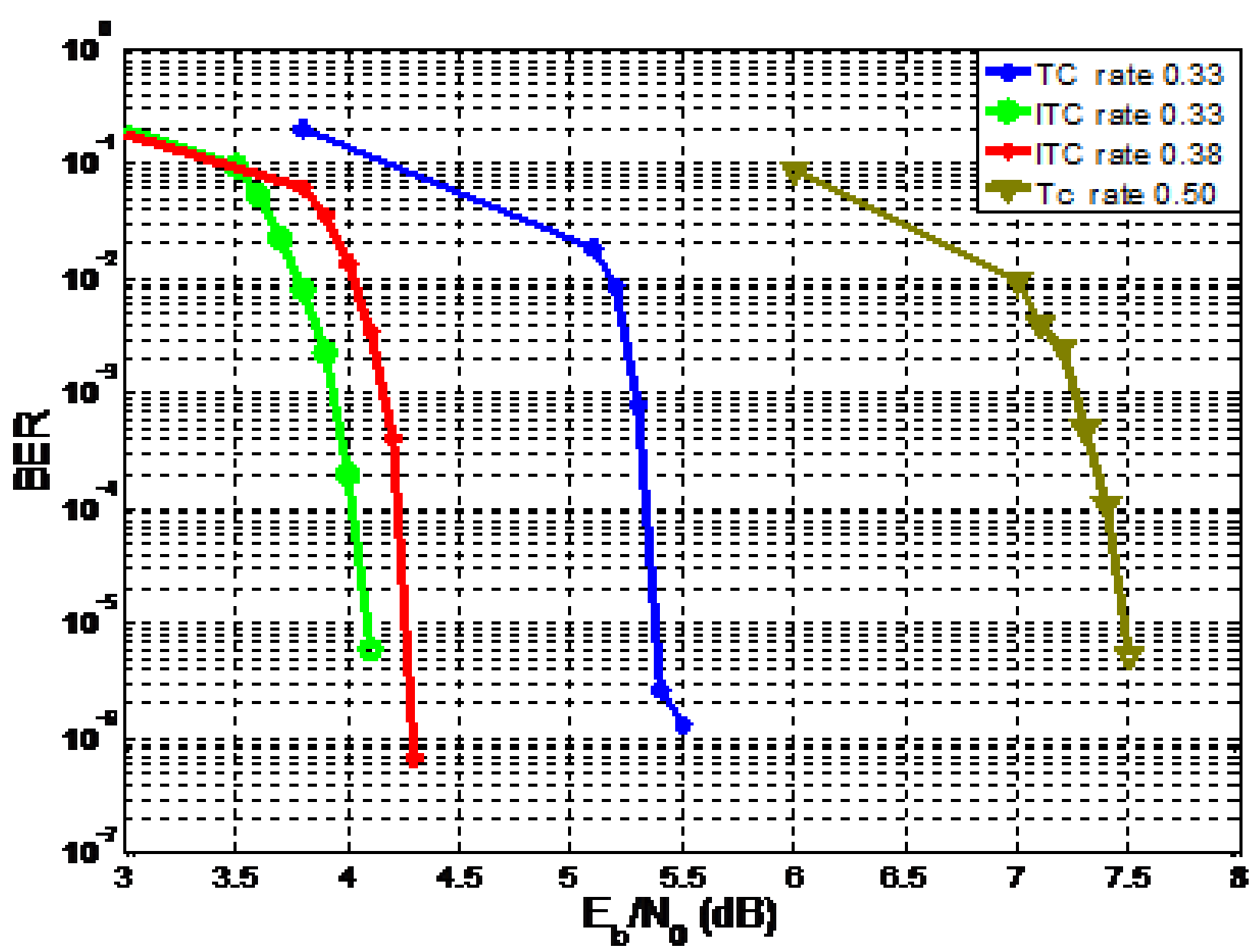

**Figure 7.** BER performance comparisons between TCs and I-TCs in an AWGN channel for BPSK modulation scheme.

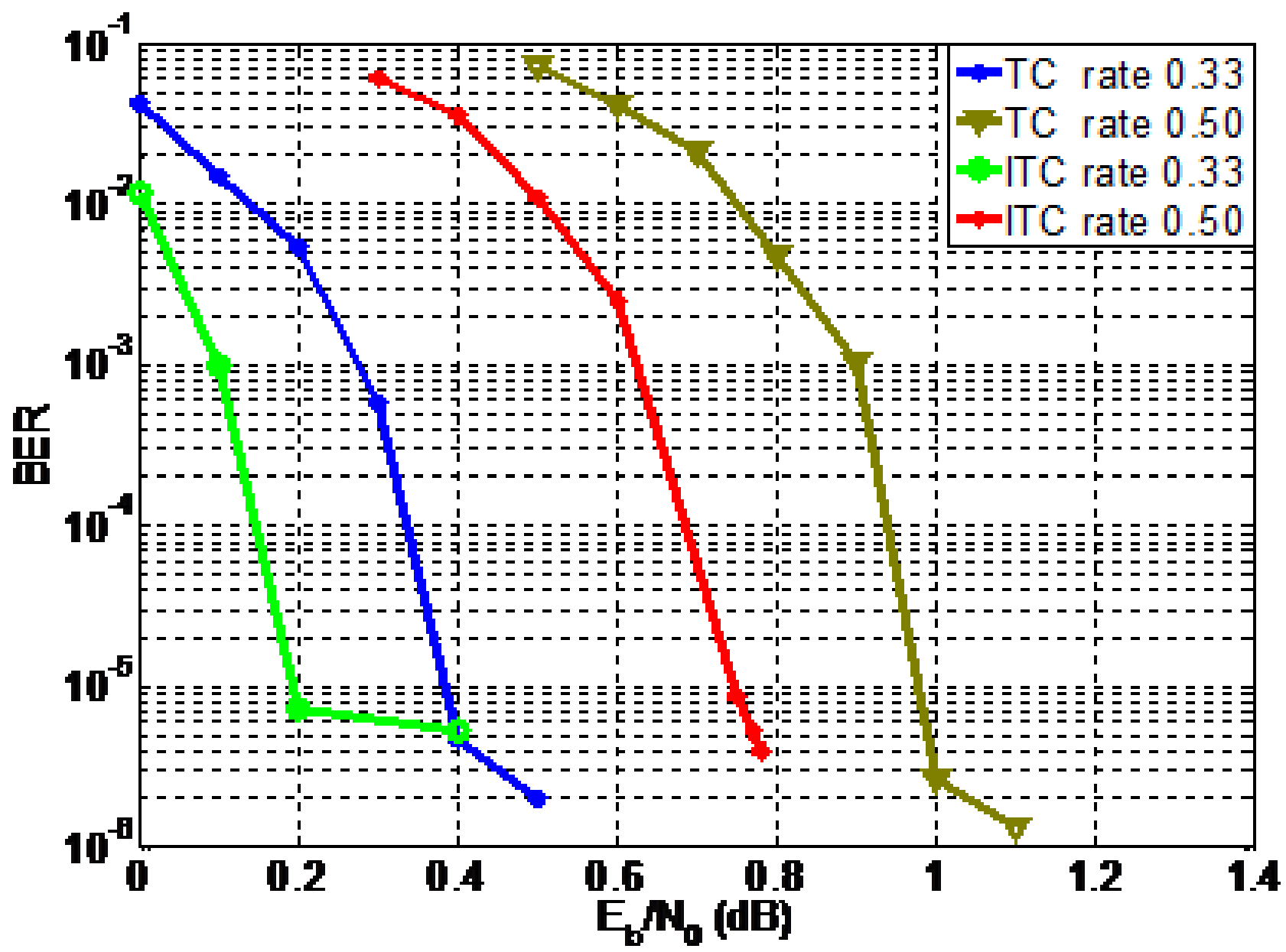

**Figure 8.** BER performance comparisons between TCs and I-TCs in an AWGN channel for QPSK modulation scheme.

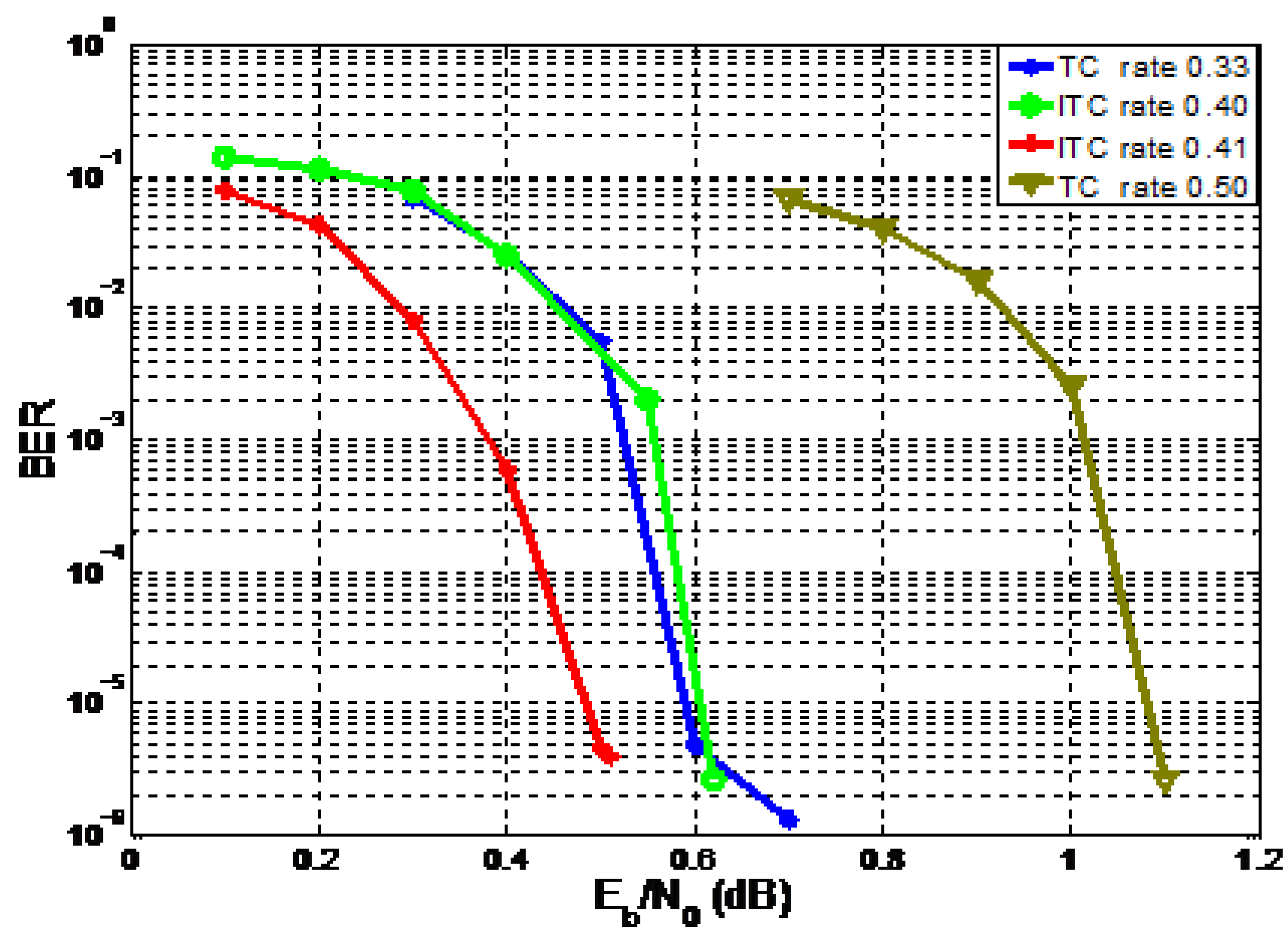

**Figure 9.** BER performance comparisons between TCs and I-TCs in an AWGN channel for 16QAM modulation scheme.

Table 2 shows the puncturing pattern, the number of iterations required to converge to a bit error rate of $10^{-5}$, the required $E_b/N_0$ value for convergence and the corresponding degree profiles used to achieve this in the I-TCs.

In the BPSK modulation scheme, the I-TC achieved a minimum coding gain of 0.2 dB over its corresponding regular TC for code rate 0.33 and 0.5 as seen in Figure 7. In the QPSK modulation scheme, the I-TC rate 0.41 has a slight advantage of about 0.09 dB coding gain over the TC rate 0.33.

Also there is a 0.07 bits per channel use increase in the I-TC over the TC at their converging $E_b/N_0$ values. Progressing to the 16QAM modulation, the I-TC rate 0.33 has a 0.3 dB coding gain over its corresponding TC rate 0.33 as seen in Figure 9.

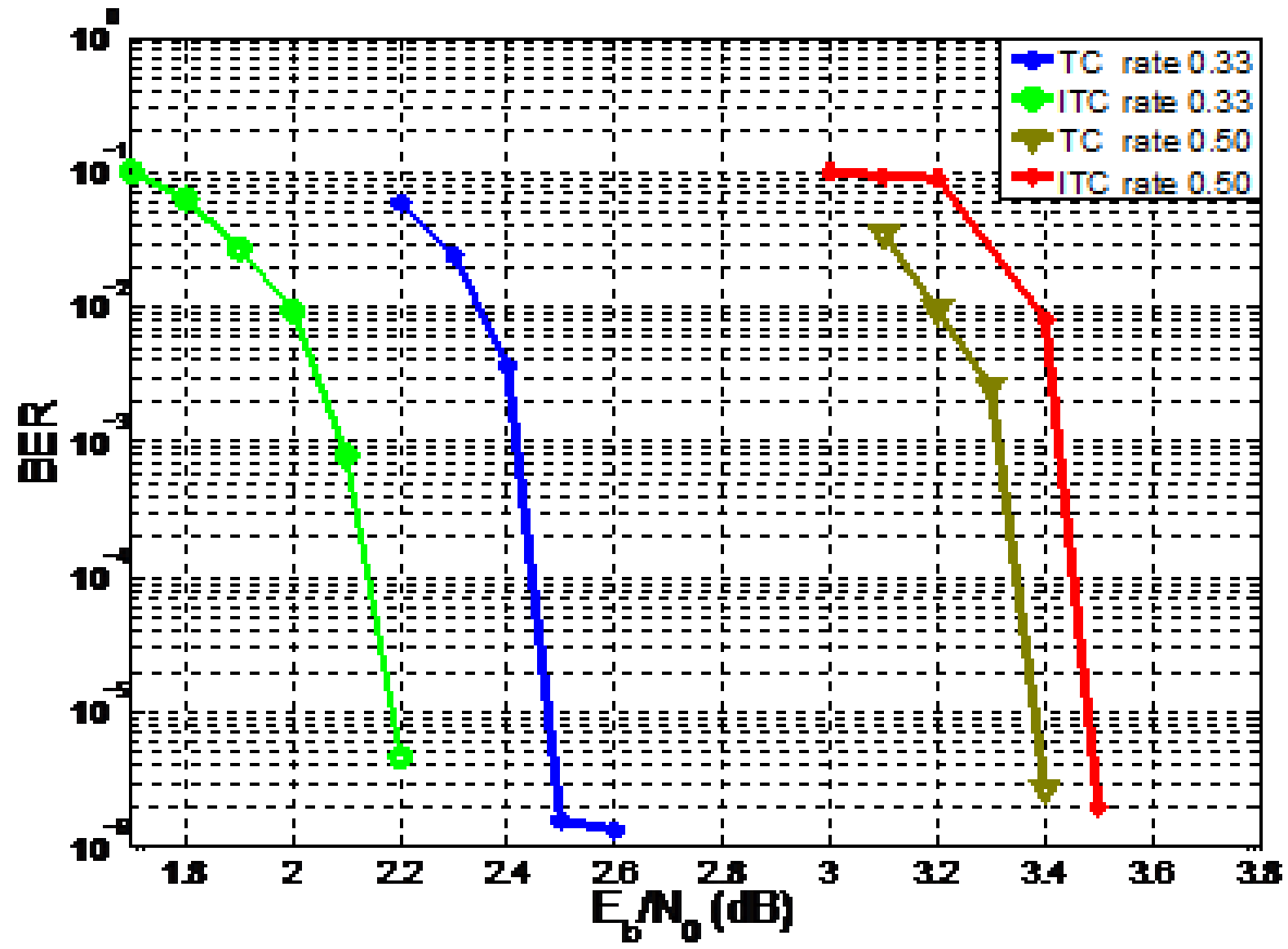

**Figure 10.** BER performance comparisons between TCs and I-TCs in AWGN for 64QAM modulation scheme.

**Table 2.** I-TC at different codes rates with their converging $E_b/N_0$ in AWGN

| Code rate | Channel | Modulation scheme | $E_b/N_0$ (dB) | $S$ | Puncture pattern | Bit degree $d_i$ (fractions $f_i$) | $I_{I-TC}$ |
|---|---|---|---|---|---|---|---|
| 0.33 | AWGN | BPSK | 0.20 | 0.32 | 11101101110 | 2(0.888), 8(0.06),9(0.052) | 31 |
| 0.50 | AWGN | BPSK | 0.75 | 0.48 | 10 | 2(0.95), 9(0.05) | 30 |
| 0.40 | AWGN | QPSK | 0.62 | 0.76 | 11101101110 | 2(0.96), 6(0.04) | 36 |
| 0.41 | AWGN | QPSK | 0.51 | 0.80 | 11101101110 | 2(0.95), 5(0.05) | 23 |
| 0.33 | AWGN | 16QAM | 2.20 | 1.28 | unpunctured | 2(0.99), 7(0.01) | 19 |
| 0.50 | AWGN | 16QAM | 3.50 | 1.96 | 10 | 2(0.94), 3(0.06) | 26 |
| 0.33 | AWGN | 64QAM | 4.10 | 1.92 | 11101101110 | 2(0.85), 7(0.15) | 16 |
| 0.38 | AWGN | 64QAM | 4.30 | 2.22 | 11101101110 | 2(0.96), 9(0.04) | 23 |

*$S$ and $I_{I-TC}$ in the table stand for the throughput in bits per channel use and the number of iterations required to converge to a low BER, respectively.

However, the TC rate 0.5 has a coding gain of 0.1 dB over the I-TC rate 0.5 at a BER of $10^{-5}$. The largest coding gain recorded in the I-TC over the TC is seen in Figure 10 i.e. the 64QAM modulation scheme, where the I-TC rate 0.33 has a significant coding gain of 1.29 dB over its corresponding TC rate 0.33. Table 3 compares the number of iterations required for a low probability of error $10^{-5}$ in a regular TC and in the I-TC in an AWGN channel using the 5012 bit frame size.

Table 3 shows that in general, the I-TC requires a higher number of iterations to converge to a low BER in comparison to the regular TC. It should be noted that the I-TC requires only one Log-MAP decoder in comparison to the regular TC which requires two Log-MAP decoders.

Figure 11 shows the throughput versus the signal power to noise power ratio for the TC and the I-TC for the AWGN channel using the 5012 bits frame size. The Shannon capacity curve shown in Figure 11 was calculated by equation (5).

$$\frac{C}{B} = \log_2\left(1+\frac{S}{N}\right) bit/s/Hz \tag{5}$$

The throughput per channel use is given by $S = R\times(\log_2 M)(1-FER)$, where $R$ is the rate of the code, $FER$ the frame error rate and $M$ the $M\!-\!ary$ order of the

**Table 3.** Number of iterations required for convergence in an I-TC and the TC

| Code rate | Modulation Scheme | $I_{TC}$ | $I_{I-TC}$ |
|---|---|---|---|
| 0.33 | BPSK | 11 | 31 |
| 0.50 | BPSK | 15 | 30 |
| 0.33 | QPSK | 15 | - |
| 0.40 | QPSK | - | 36 |
| 0.41 | QPSK | - | 23 |
| 0.50 | QPSK | 11 | - |
| 0.33 | 16QAM | 14 | 19 |
| 0.50 | 16QAM | 21 | - |
| 0.33 | 64QAM | 24 | 16 |
| 0.38 | 64QAM | - | 23 |
| 0.50 | 64QAM | 19 | - |

* $I_{TC}$ and $I_{I-TC}$ in the table stand for the number of iterations required to converge to a low BER in the regular TC and the I-TC, respectively.

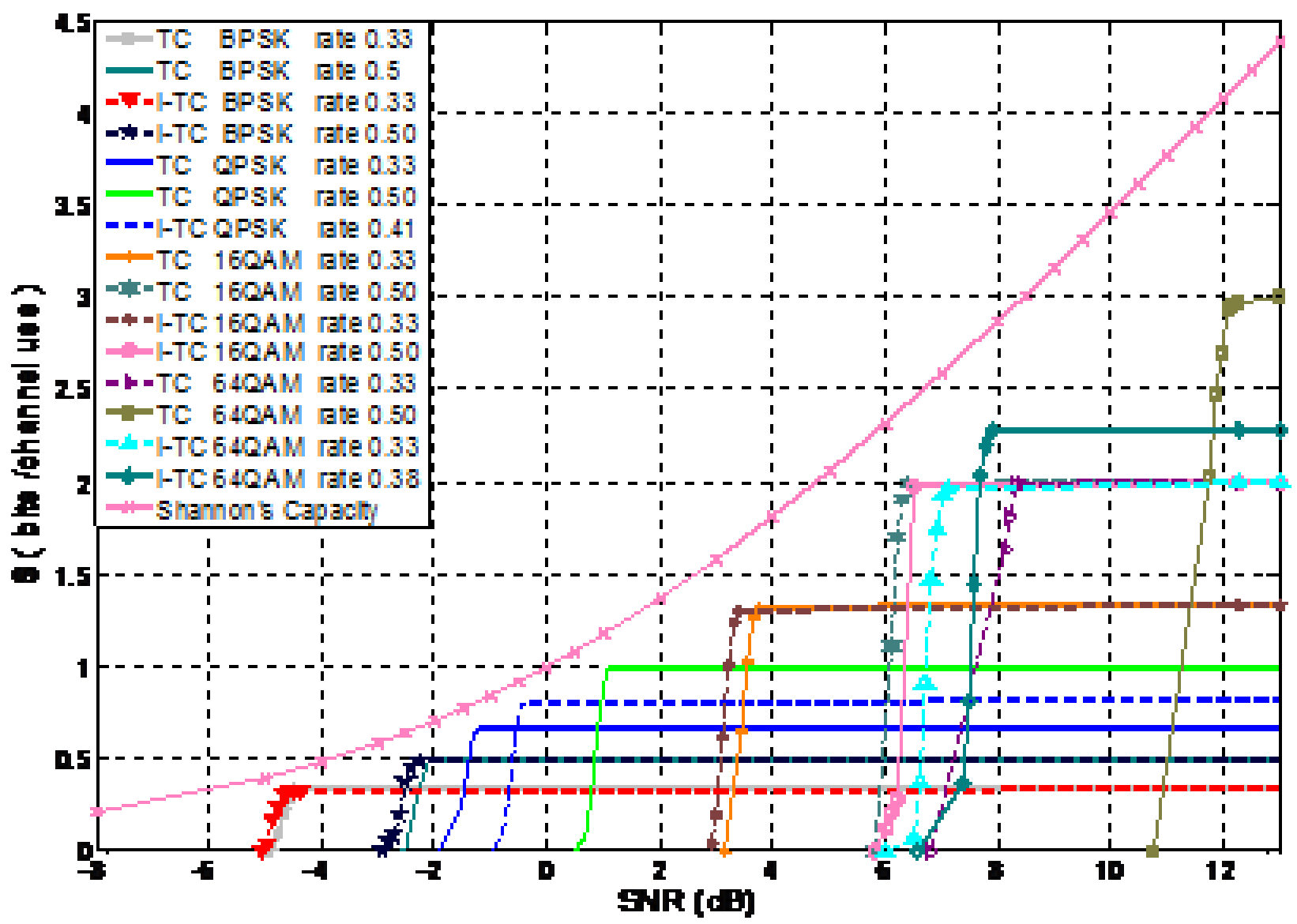

**Figure 11.** Throughput versus SNR for TCs and I-TCs in an AWGN channel for different modulation schemes.

modulation scheme. The frame error rate is computed by dividing the number of frames in error after decoding by the total number of frames used in the simulation.

The frame size used is 5012 bits. The SNR difference between Shannon's capacity and the throughput curves is smaller for the I-TCs in all the modulation schemes and code rates considered than the TCs, except for the 16QAM code rate 0.5 where the I-TC and the TC are of equal distance to Shannon's capacity.

The behavioral pattern of the throughput curve for the I-TC and the TC are very similar with the exception that the I-TCs have higher throughput values. An exception to this is in the case of the 64QAM code rate 0.50 where the TC has a higher throughput value but at a significant 4.1 dB loss in comparison to the I-TC rate 0.38. In general, the I-TC gives a better performance in comparison to the TC in terms of throughput value at a given SNR value and does not give a worse performance.

## 5. Conclusion

In this paper, the design and construction of a high performing I-TC has been shown. The BER versus Eb/N0 performance of the I-TC in comparison to the regular TC using different frame sizes were also shown where larger frame sizes, had larger coding gains in the I-TCs and the TCs.

Furthermore, the I-TC in this paper has been shown to be capable of converging to a low probability of error with lesser complexity in comparison to other I-TCs in terms of the number of iterations and frame sizes.

In comparison to the regular TC, the I-TC in general uses a single recursive Convolutional Code and a single soft-input soft-output decoder. Also, the designed I-TC performs no worse than the regular TC and frequently better, for example, it is capable of achieving a coding gain of 1.29 dB over its corresponding regular TC, when used in an AWGN channel with 64QAM modulation making it an advantageous alternative to the regular TC in future communication systems.